\documentclass[pra,twocolumn,showpacs]{revtex4}
\usepackage[koi8-r]{inputenc}
\usepackage[english]{babel}
\usepackage{amsmath}
\usepackage{amssymb}
\usepackage{epsfig}

\begin{document}
\title{\bf Quasi-stable configurations of torus vortex knots and links}
\author{Victor~P. Ruban}
\email{ruban@itp.ac.ru}
\affiliation{L.D. Landau Institute for Theoretical Physics RAS, 
142432 Chernogolovka, Moscow region, Russia} 

\date{\today}

\begin{abstract}
The dynamics of torus vortex configurations $V_{n,p,q}$ in a superfluid liquid 
at zero temperature ($n$ is the number of quantum vortices, $p$ is the number 
of turns of each filament around the symmetry axis of the torus, 
and $q$ is the number of turns of the filament around its central circle; 
radii $R_0$ and $r_0$ of the torus at the initial instant are much larger 
than vortex core width $\xi$) has been simulated numerically based on a 
regularized Biot-Savart law. The lifetime of vortex systems till the instant 
of their substantial deformation has been calculated with a small step in 
parameter $B_0=r_0/R_0$ for various values of parameter $\Lambda=\log(R_0/\xi)$. 
It turns out that for certain values of $n$, $p$, and $q$, there exist 
quasi-stability regions in the plane of parameters $(B_0,\Lambda)$, 
in which the vortices remain almost invariable during dozens and 
even hundreds of characteristic times.
\end{abstract}
\pacs{47.32.C-, 47.37.+q, 67.25.dk}
\maketitle

\section{Introduction}

Vortex knots and links have been objects of
interest in the classical fluid dynamics since the 19-th
century. In particular, Lord Kelvin \cite{Kelvin} put forth the
hypothesis that uniformly rotating and propagating
(along a certain axis) symmetric stationary configurations 
of $n$ thin vortex filaments (each with circulation $\Gamma$)
can exist in an ideal fluid. The shape of the filaments 
is close to torus vortices $V_{n,p,q}$ that are determined 
parametrically by the following expressions:
\begin{eqnarray}
\label{X_tor}
X_j(\beta)&=&\Big[R_0+r_0\sin\Big(q\beta+\frac{2\pi j}{np} 
-\Theta_0\Big)\Big]\cos(p\beta),\\
\label{Y_tor}
Y_j(\beta)&=&\Big[R_0+r_0\sin\Big(q\beta+\frac{2\pi j}{np} 
-\Theta_0\Big)\Big]\sin(p\beta),\\
\label{Z_tor}
Z_j(\beta)&=&r_0\cos\Big(q\beta+\frac{2\pi j}{np} -\Theta_0\Big) +Z_0,
\end{eqnarray}
where $j=1,\dots, n$  is a vortex number, $n$ is the number
of quantized vortex filaments, longitudinal parameter
$\beta$ runs through the interval $0\le\beta <2\pi$, and $\Theta_0$ and $Z_0$
are linear functions of time $t$. Natural numbers $p$ and $q$
must have no common multipliers. It can easily be
seen that $p$ is the number of turns of each filament around
the symmetry axis of the torus and $q$ is the number of
turns around its central circle. If $p > 1$ and $q > 1$, each
vortex filament is a torus knot ${\cal T}_{p,q}$. If at least one of
these numbers ($p$ or $q$) is equal to unity, each line is an
unknot,  ${\cal U}_{p,1}$ or ${\cal U}_{1,q}$, but it is linked in this case
with all remaining lines (for $n\ge 2$). If $p=1$ and $q=1$,
we obtain a link of $n$ rings.

Despite such a long history of the problem, such
vortex knots and links have been produced experimentally 
for the first time quite recently \cite{KI2013}.

It should be noted that real objects closest to the
theory are quantized vortex filaments in superfluid
liquids (e.g., helium at a low temperature, for which
the effect of the normal component is negligibly small).
In this case, $\Gamma=2\pi\hbar/m_{\rm at}$ is the velocity circulation
quantum ($m_{\rm at}$ is the mass of the helium atom), and the
width of the core of each vortex is $\xi$. Unfortunately,
the experimental technique for producing knots and
links for quantum vortices has not been developed as yet.

It should be observed that exact stationary configurations
of the torus type have not been found (even
numerically) as yet, and their stability has not been
tested. The difficulties are associated to a considerable
extent with the fact that the steady-state solutions correspond 
not to the energy functional minimum for preset values of the momentum 
and angular momentum, but only to a saddle point. Therefore, we have
to operate with approximate formulas (1)--(3). In a
number of publications, the dynamics of torus vortices
was simulated numerically for a small number of sets
of parameters and for not very long time intervals,
during which vortices could propagate along $z$ axis
by not more than a few dozen $R_0$ ($R_0$ and $r_0$ are the radii
of the torus), and then the deformation of the filaments
increased [3-11]. In the quantum-mechanical case,
this led to reconnections. It may appear that such
results indicate an (albeit relatively weak) instability of
torus knots and links.

The actual situation is more complicated and interesting. 
In the author's recent publication [12], it was
shown for simplest knots ${\cal T}_{2,3}$ and ${\cal T}_{3,2}$ that in the
space of parameters $B_0=r_0/R_0$ and $\Lambda=\log(R_0/\xi)$
($\xi$ is the vortex core width), there exist quasi-stability
regions that are gaps between fundamental parametric
resonances of various types, where a knot remains
almost unchanged on the average for many dozen and
even hundreds of characteristic times, passing distances 
that are sometimes equal to thousands of initial
radii $R_0$. In other words, the toric shape of a vortex
contains perturbation modes relative to the corresponding 
(unknown) stationary configuration, and the amplitudes 
of these modes increase with time in not all cases.

To determine quasi-stable regions, computer calculations 
of the lifetime of knots till the instant of their
significant deformation with a small step in parameter $B_0$ 
for a preset value of $\Lambda$ were required. All quasi-stable
zones of a trefoil knot ${\cal T}_{2,3}$ determined in [12] have
a small width $\Delta B_0\lesssim 0.01$ and correspond to relatively
``thin'' tori with $B_0\lesssim 0.2$ (see the top panel in Fig. 1
below). It is significant that the maximal values of $B_0$
are attained for $\Lambda\approx 3.5$. For $\Lambda\lesssim 3$, such zones are
absent, while for $\Lambda\gtrsim 6$, the zones are shifted to small
$B_0\lesssim 0.1$. These regions have not been observed earlier
in all probability precisely because of their ``periphery'' position and a small size.

A natural question arises: what is the situation with
other knots and links? This study aims at the search
for analogous quasi-stable configurations also for other
$V_{n,p,q}$ apart from $V_{1,2,3}$ and $V_{1,3,2}$. It will be shown below
that such configurations exist for not all sets $\{n,p,q\}$ (at
least in the range of $B_0\gtrsim 0.1$ of interest; for small values
of $n$, this range corresponds to considerable distances $l\sim 2r_0\gtrsim 4\xi$
 between vortex cores for $\Lambda\approx 3$).

\begin{figure}
\begin{center}
\epsfig{file=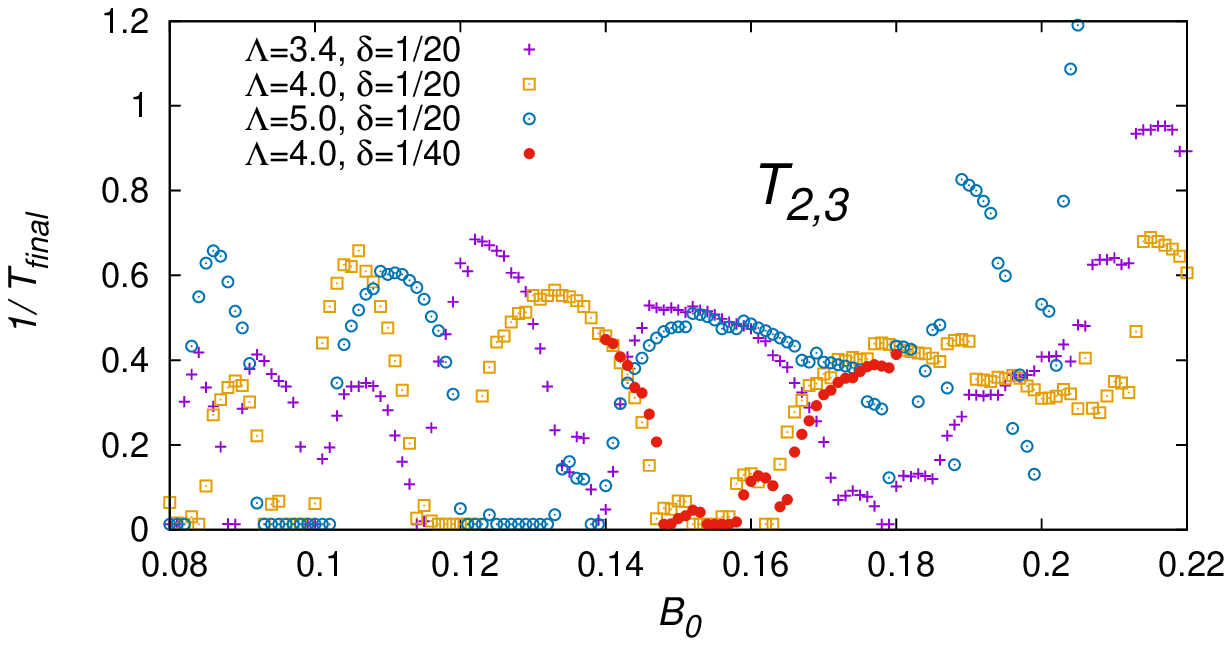, width=82mm}\\
\epsfig{file=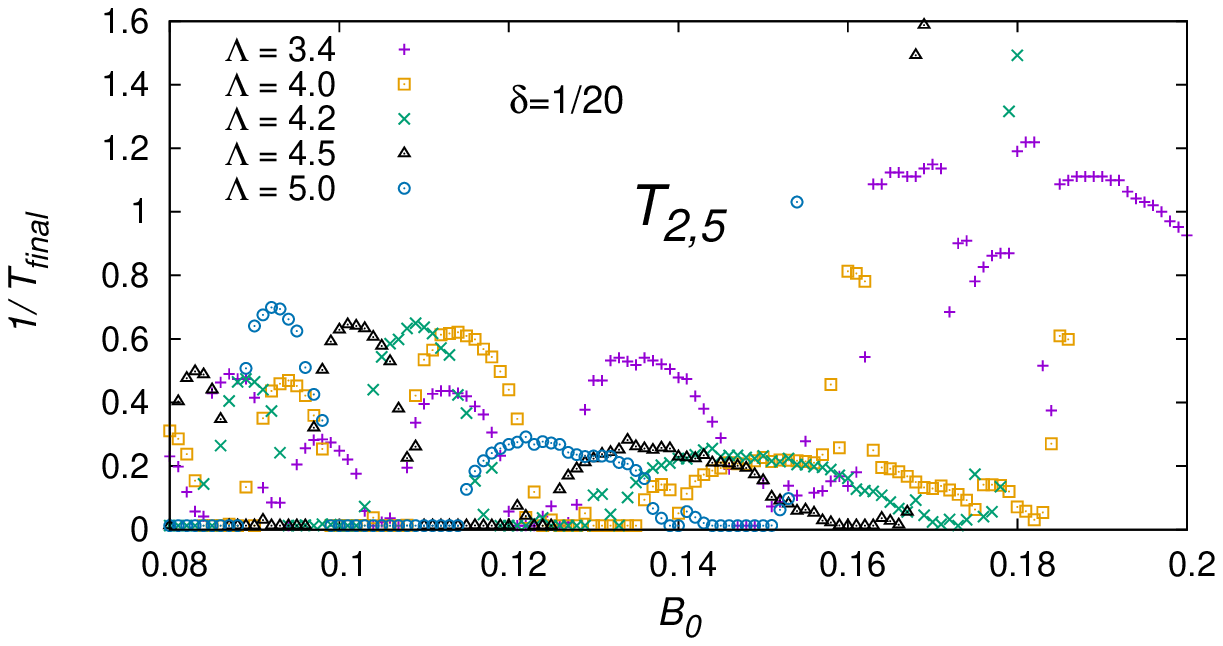, width=82mm}\\
\epsfig{file=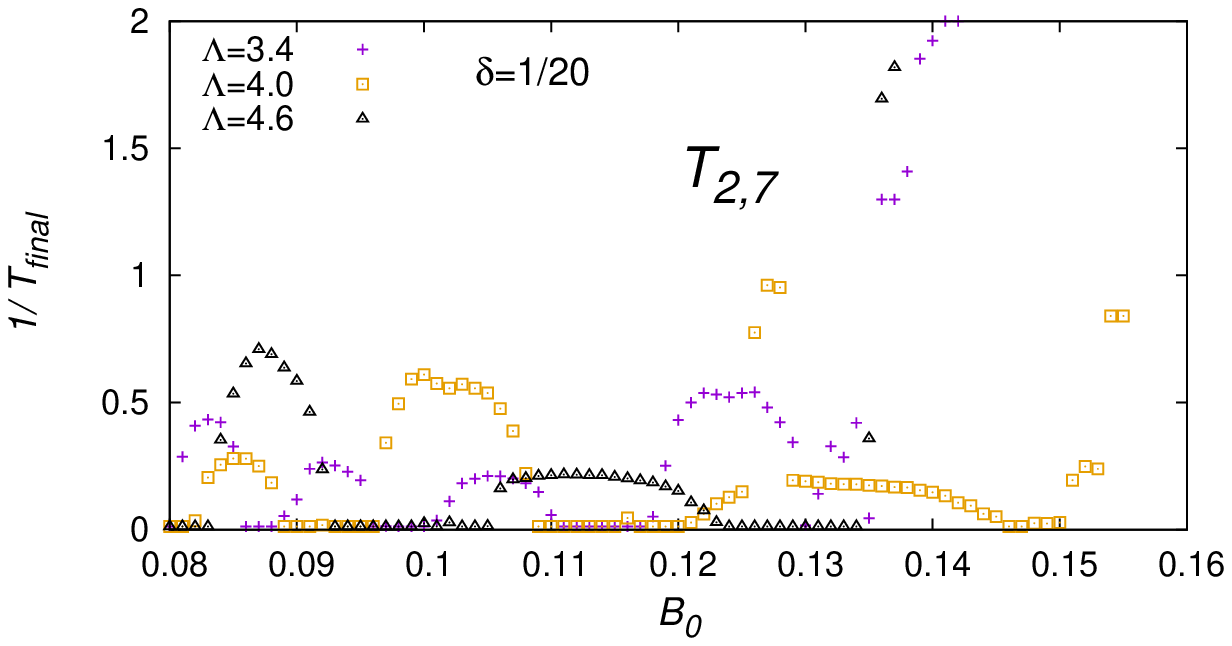, width=82mm}
\end{center}
\caption{Reciprocal lifetime of vortex knots ${\cal T}_{2,q}$ for $q=3,5,7$
for different parameters $\Lambda$ and $B_0$. In all
three panels, the quasi-stability zones in the form of segments 
close to the horizontal axis can be seen.}
\label{T2q} 
\end{figure}
\begin{figure}
\begin{center}
\epsfig{file=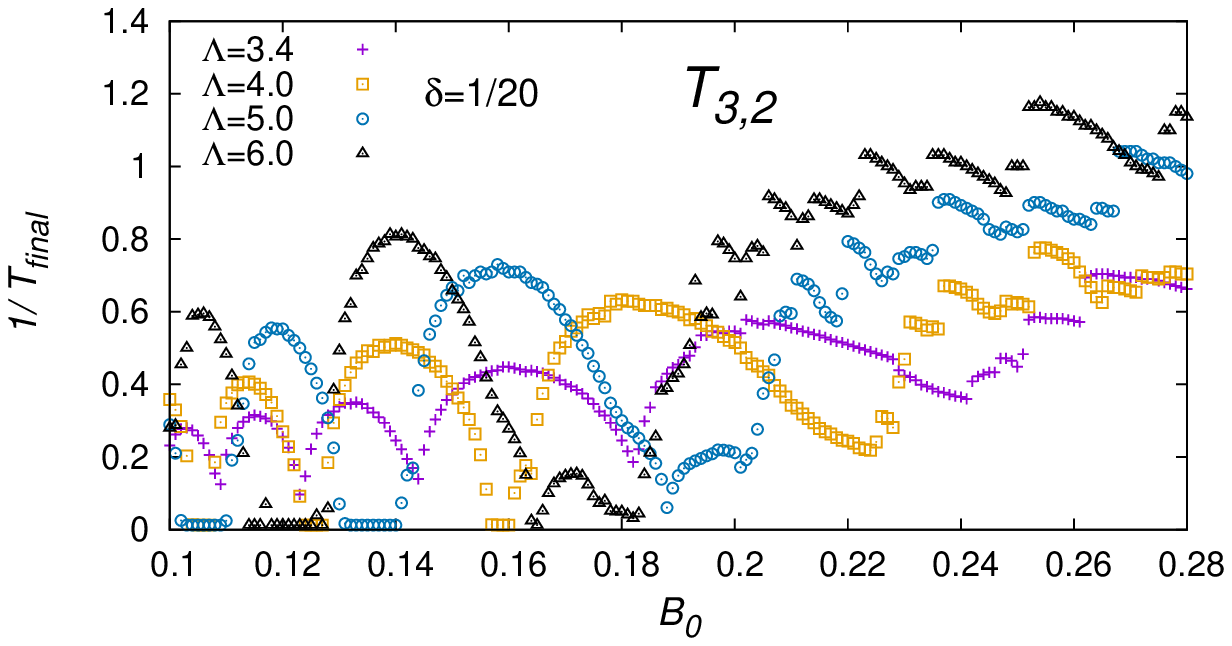, width=82mm}\\
\epsfig{file=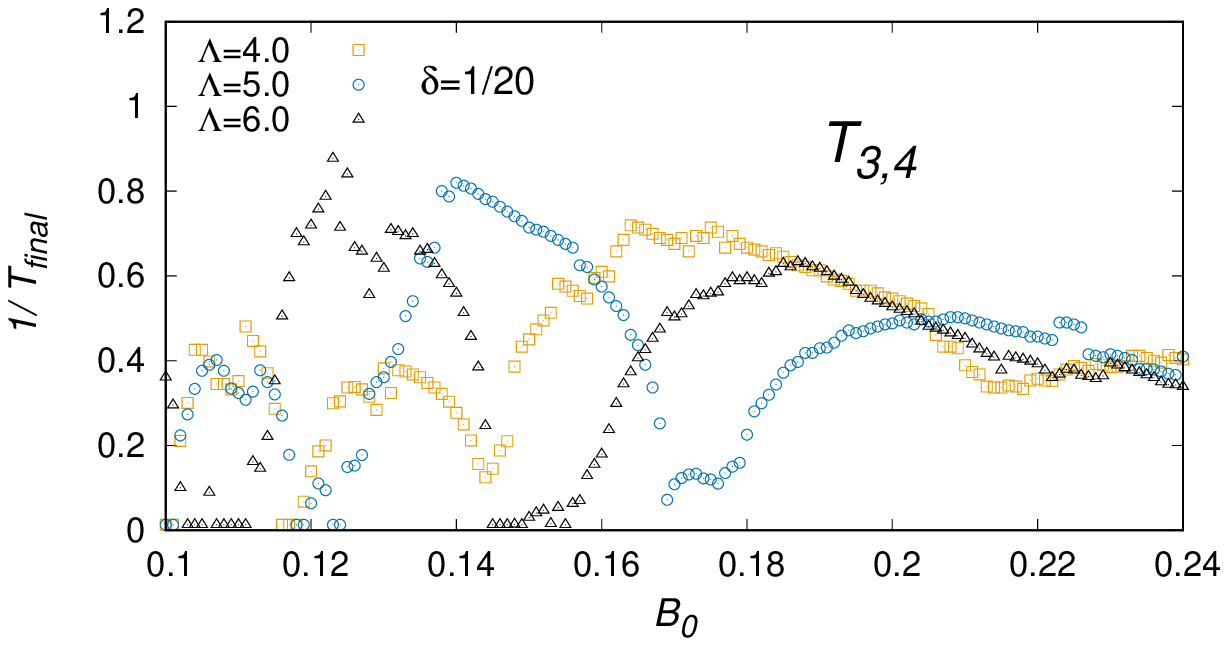, width=82mm}\\
\epsfig{file=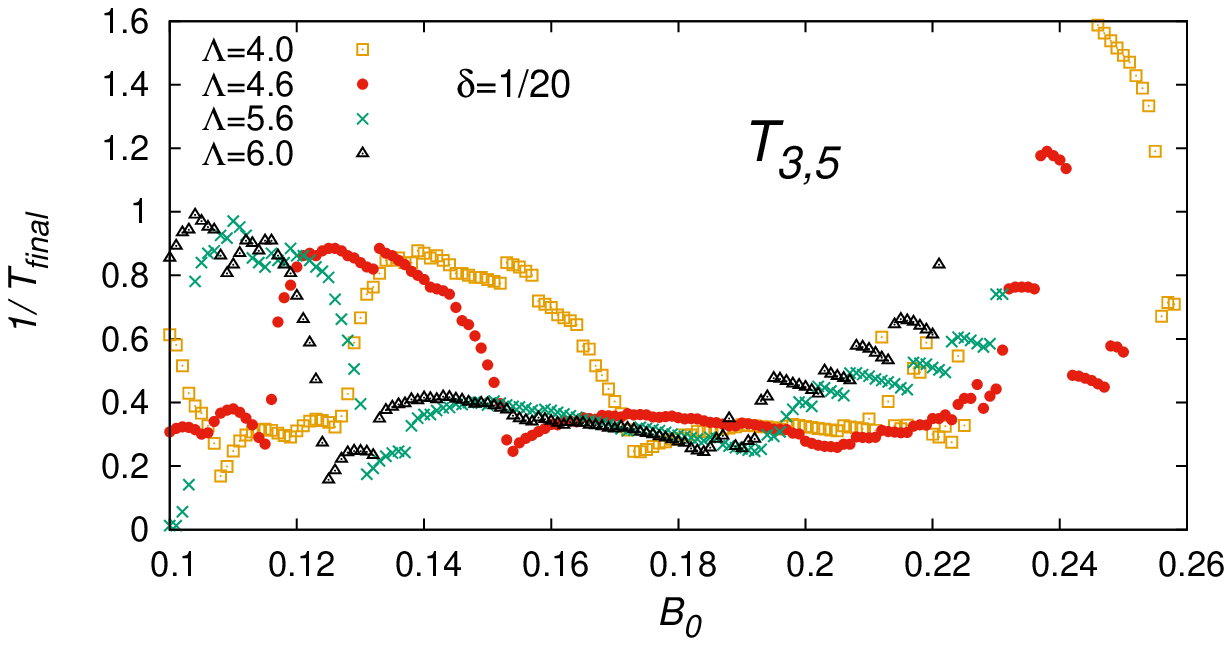, width=82mm}\\
\epsfig{file=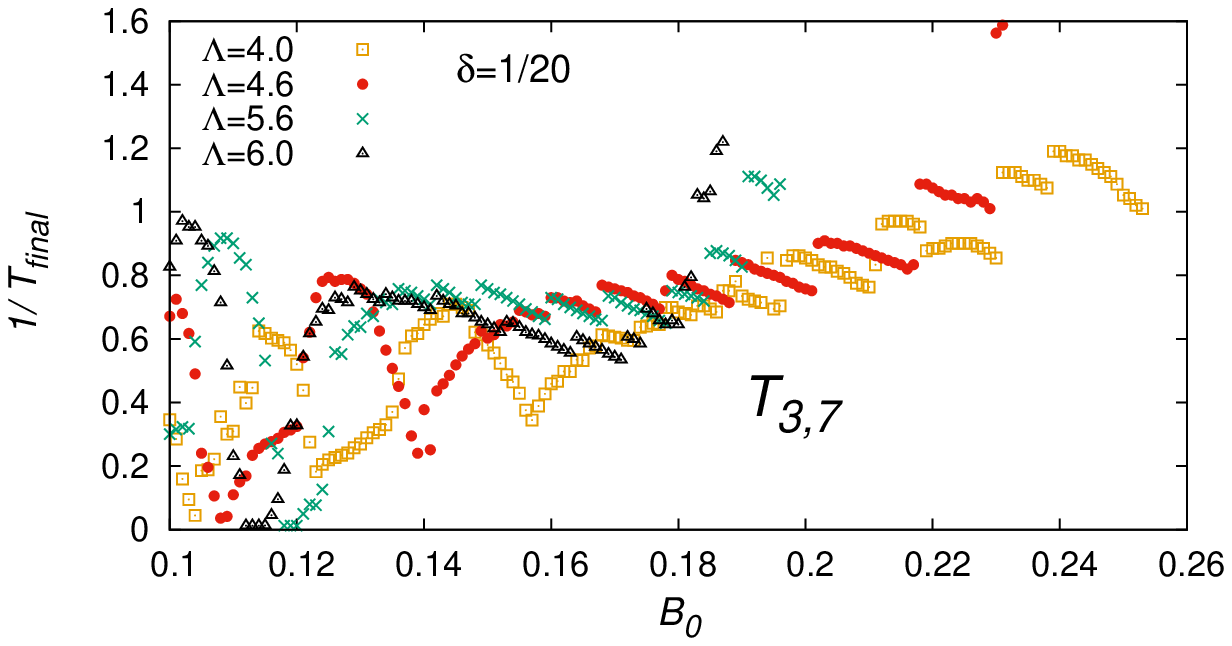, width=82mm}
\end{center}
\caption{Reciprocal lifetime of vortex knots ${\cal T}_{3,q}$ for $q=2,4,5,7$.
The quasi-stability zones are manifested most clearly for $q=2$  and $4$.}
\label{T3q} 
\end{figure}
\begin{figure}
\begin{center}
\epsfig{file=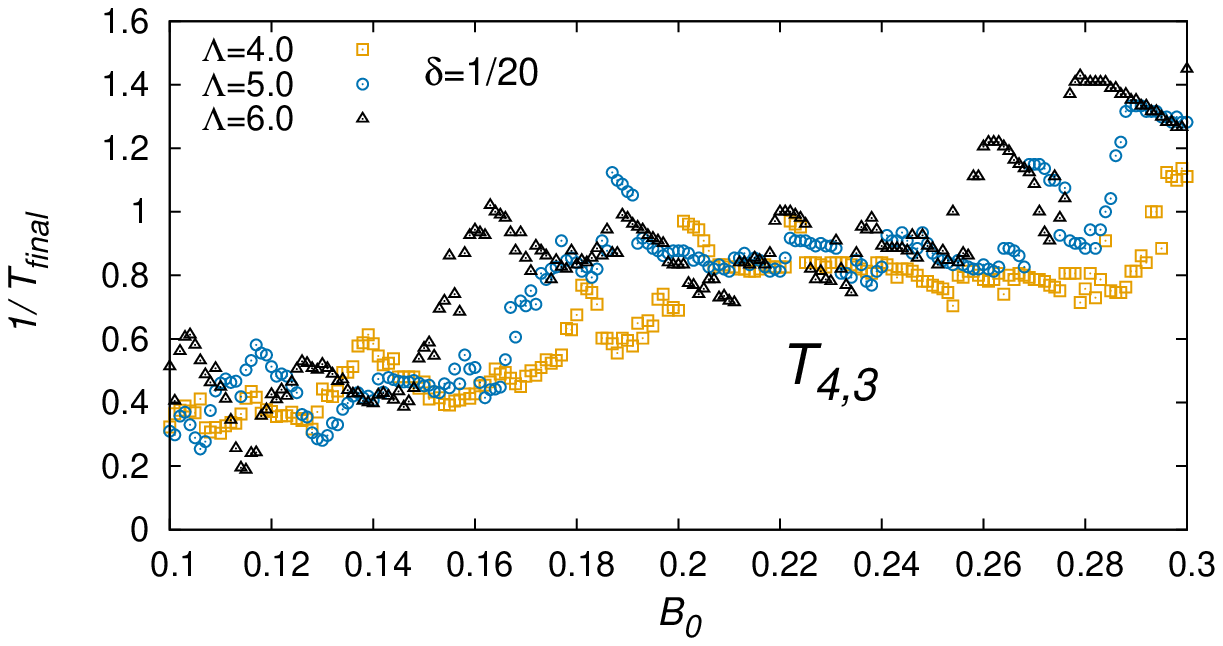, width=82mm}\\
\epsfig{file=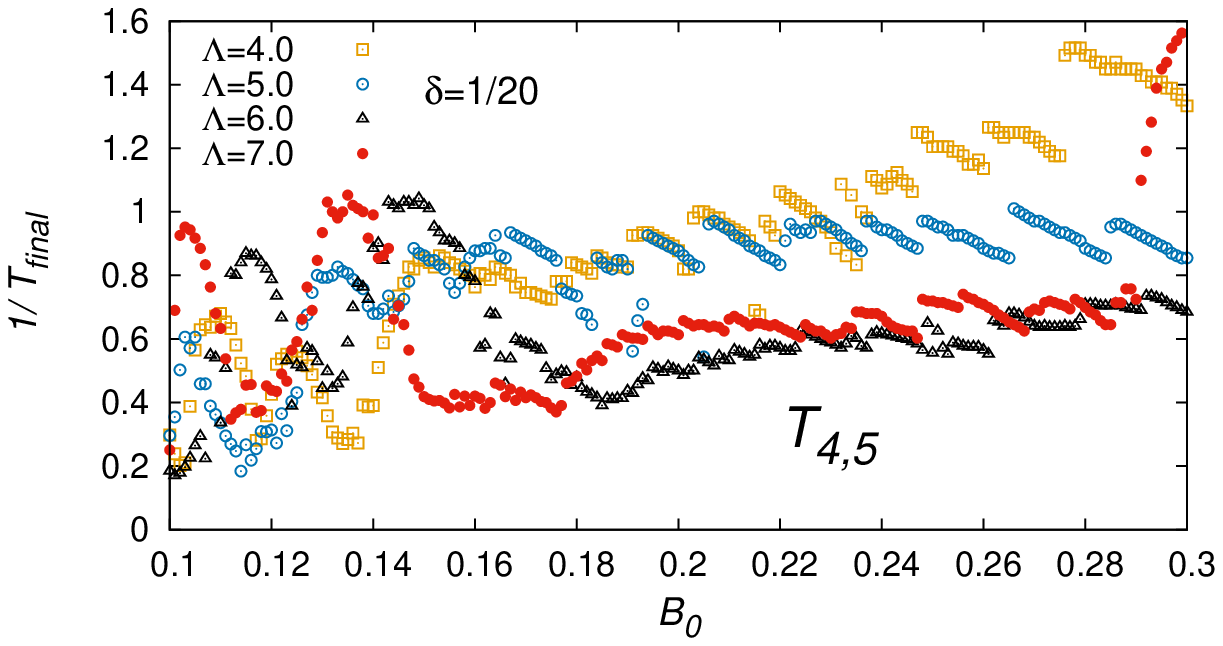, width=82mm}
\end{center}
\caption{Reciprocal lifetime of vortex knots ${\cal T}_{4,3}$ and ${\cal T}_{4,5}$. 
The quasi-stability zones are absent.}
\label{T43T45} 
\end{figure}
\begin{figure}
\begin{center}
\epsfig{file=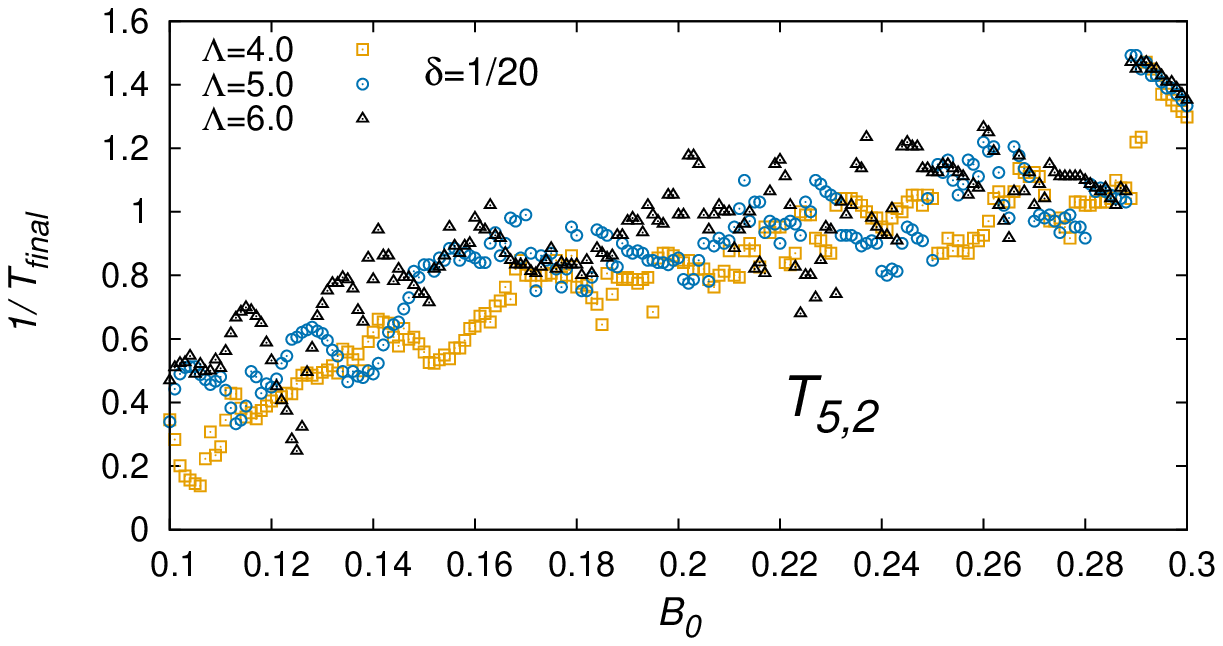, width=82mm}\\
\epsfig{file=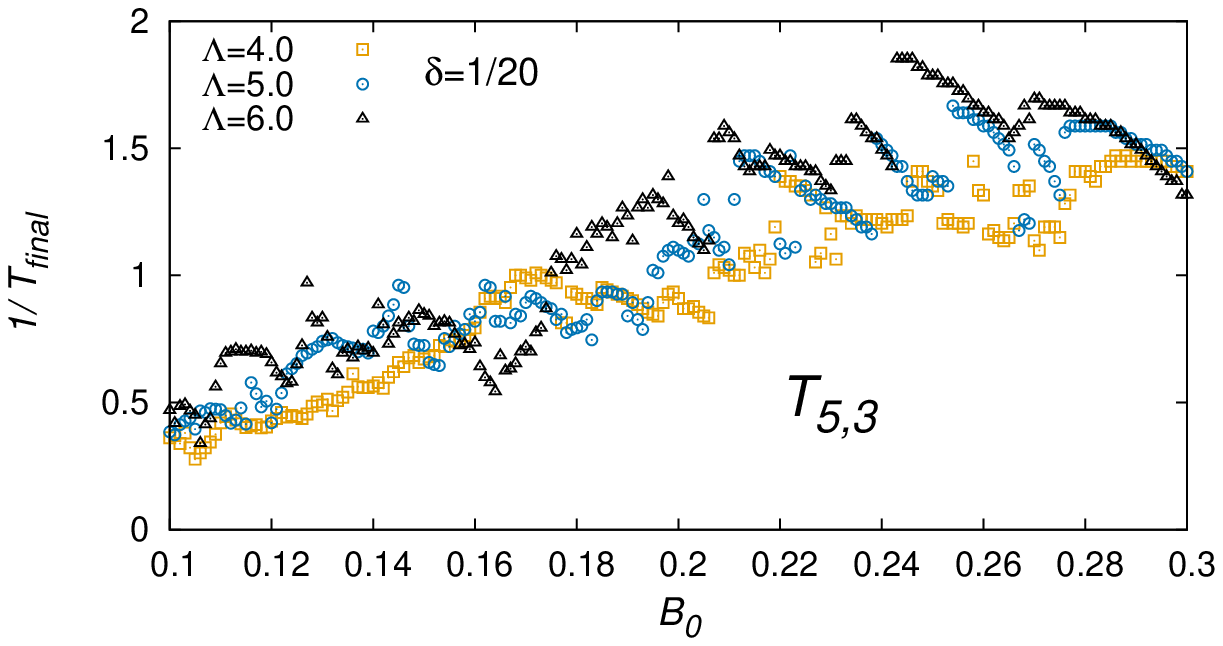, width=82mm}\\
\epsfig{file=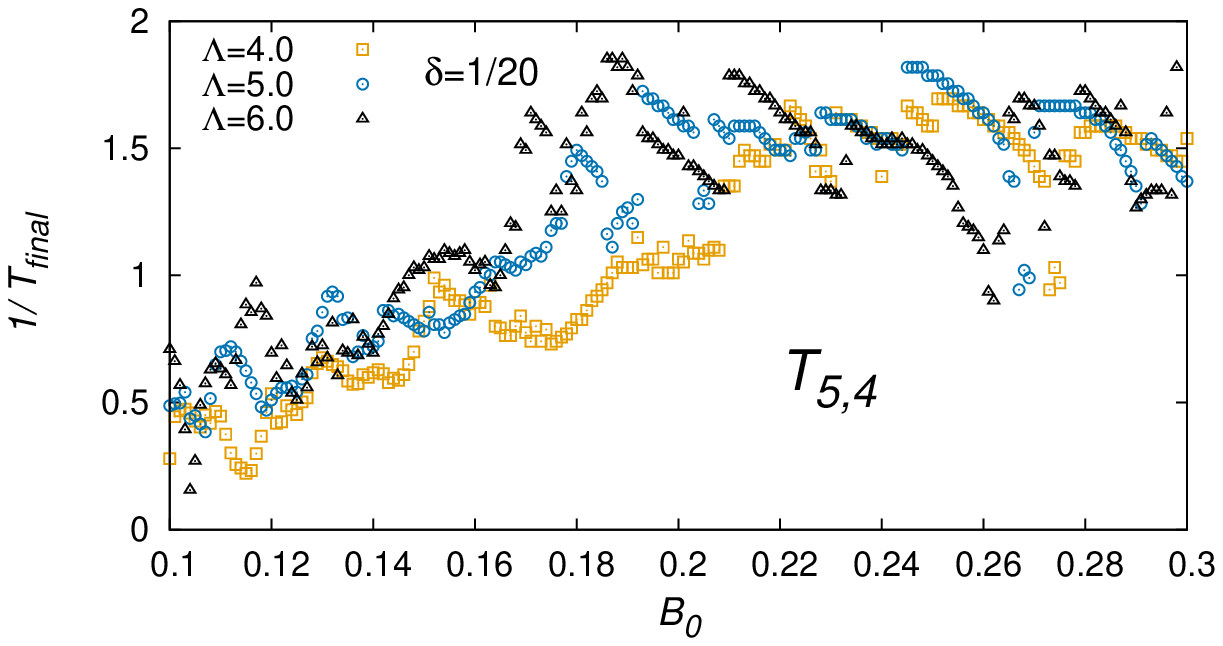, width=82mm}
\end{center}
\caption{Reciprocal lifetime of vortex knots ${\cal T}_{5,q}$ for $q=2,3,4$. 
The quasi-stability zones are absent.}
\label{T52T53} 
\end{figure}

\begin{figure}
\begin{center}
\epsfig{file=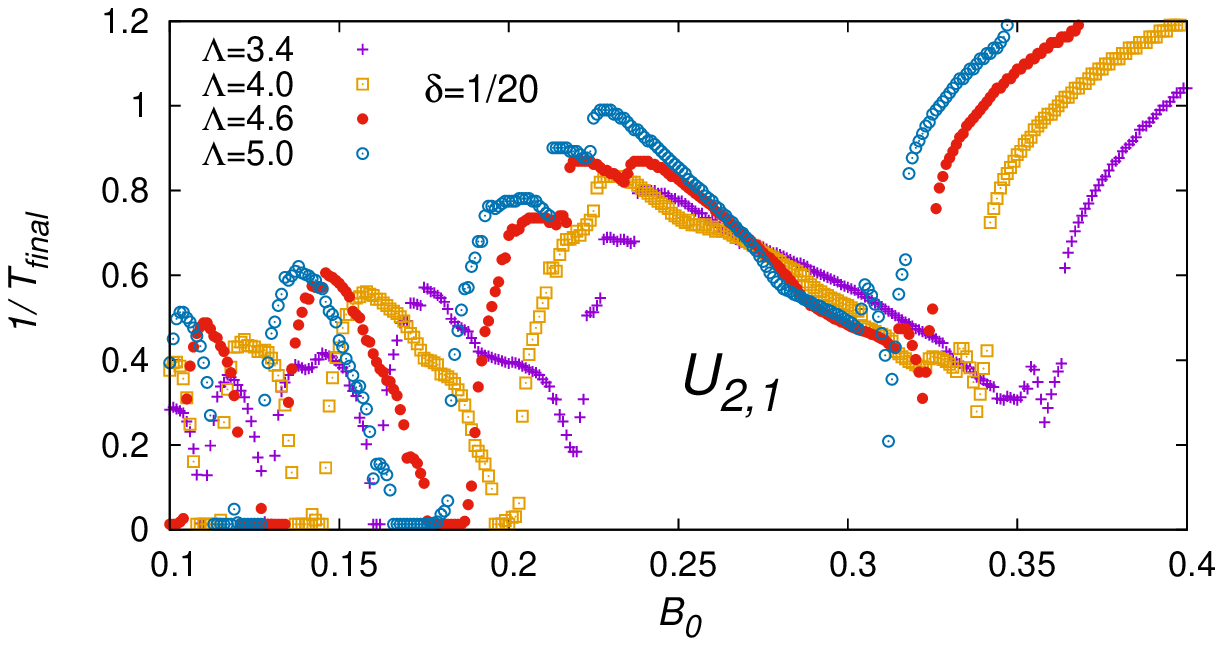, width=82mm}\\
\epsfig{file=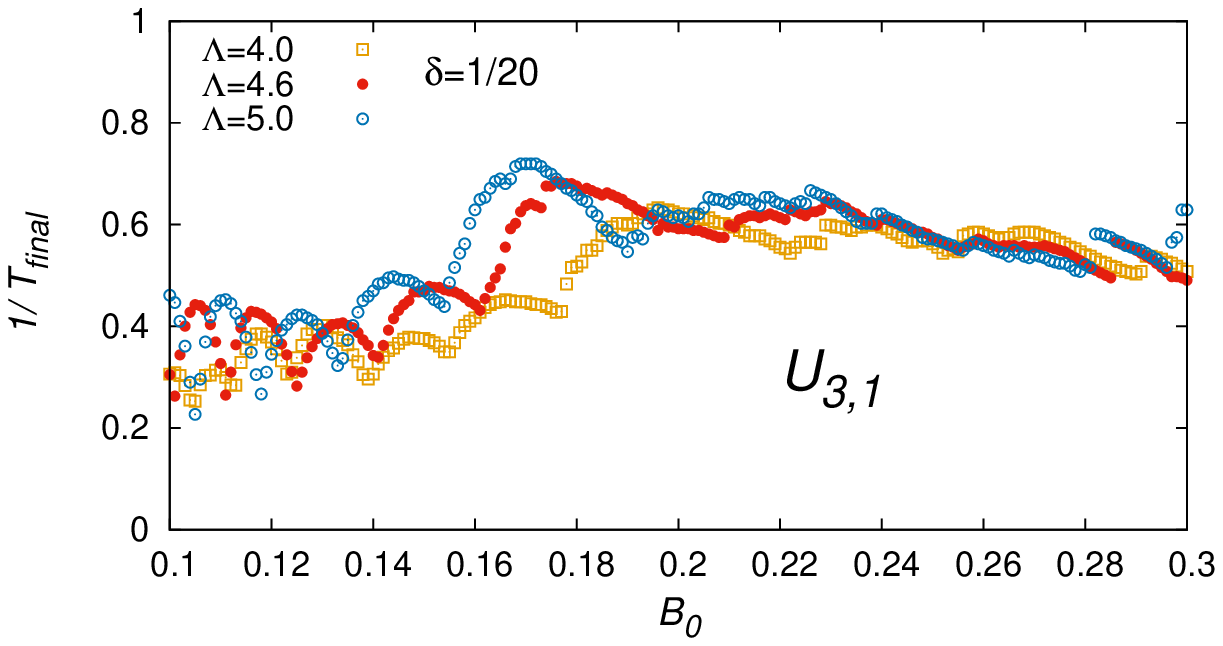, width=82mm}\\
\epsfig{file=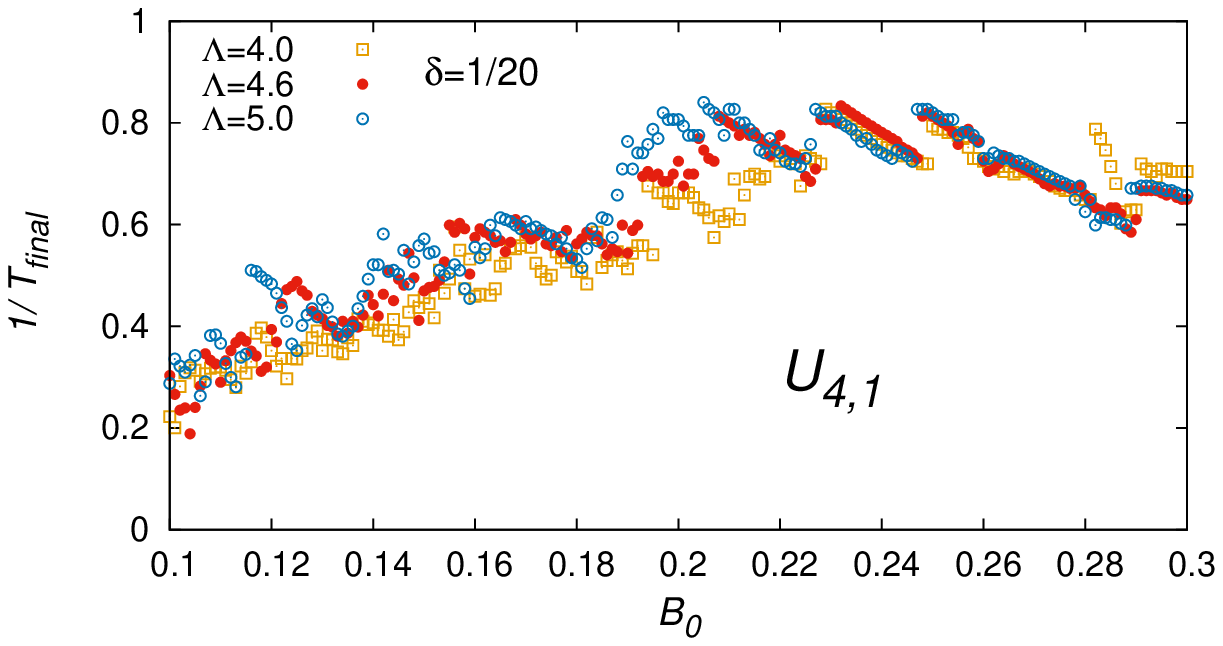, width=82mm}
\end{center}
\caption{Reciprocal lifetime of vortex unknots 
${\cal U}_{2,1}$, ${\cal U}_{3,1}$, and ${\cal U}_{4,1}$. For ${\cal U}_{2,1}$, 
quasi-stable configurations exist even for large $B_0\approx 0.2$.}
\label{U21U31U41} 
\end{figure}

\section{Dynamic model}

Let us recall that the dynamics of several thin vortex 
filaments is governed to a high degree of accuracy
by the regularized Biot-Savart law combined with the
local induction contribution (see, for example, [13-15] 
and the vast literature cited therein),
\begin{equation}
\dot{\bf X}_j(\beta,t)=\sum_{i=1}^{n}\frac{\Gamma}{4\pi}
\oint \frac{\tilde{\bf X}_i'\times({\bf X}_j-\tilde{\bf X}_i)}
{\mbox{reg}_\xi|{\bf X}_j-\tilde{\bf X}_i|^3}d\tilde\beta
+\frac{\Gamma\Lambda_0}{4\pi}\varkappa_j{\bf b}_j,
\label{BS_reg}
\end{equation}
where $\tilde{\bf X}_i={\bf X}_i(\tilde\beta,t)$, 
$\tilde{\bf X}_i'=\partial {\bf X}_i(\tilde\beta,t)/\partial\tilde\beta$, 
$\Lambda_0$ is a dimensionless positive parameter of the order of unity, which
characterizes the vortex core; $\varkappa_j$ is the local curvature
of the $j$-th filament, and ${\bf b}_j$ is the local unit binormal
vector. The method of regularization of logarithmically diverging integrals 
almost does not affect the dynamics of the filament if parameter $\Lambda_0$  
is defined concordantly. For this purpose, the Rosenhead-Moore approximation 
is often used: 
\begin{equation}
\mbox{reg}_\xi|{\bf X}_2-{\bf X}_1|^3=\sqrt{(|{\bf X}_2-{\bf X}_1|^2+\xi^2)^3}.
\label{reg}
\end{equation}

Obviously, the dynamics of vortex filaments as 1D
objects in the 3D space is invariant to arbitrary (regular) 
replacements of longitudinal parameter $\beta$. This
fact makes it possible to supplement the right-hand
sides of equations of motion (4) with terms of the form
$\mu_j{\bf X}'_j/|{\bf X}'_j|$ with arbitrary functions $\mu_j$. 
These functions can be chosen so that uncontrollable excessive 
crowding or sparseness of discrete points approximating the
vortex line is avoided. For example, we can take $\mu_j$  in
the form $\mu_j=C|{\bf X}'_j|'$, which ensures the favorable tendency 
to the uniform distribution of points along the curve. 
Another approach (to choose $\mu_j$ so that the azimuthal 
component of vector $\dot{\bf X}_j$ vanishes) is suitable
for situations when the geometric center of the perturbed 
torus vortex structure does not move far from the $z$ axis with time. 
In this study, both variants are used.

It is also important for applications that the long-wavelength dynamics 
of system (4) is just weakly sensitive to the change of the parameters
\begin{equation}
\xi\to\delta, \qquad \Lambda_0\to\Lambda_0+\log(\delta/\xi),
\end{equation}
where $\delta$ is an arbitrary quantity on the order of $\xi$ if only
the configuration of filaments is quite far from intersections. 
In particular, we can redefine parameter $\xi$  so
that $\Lambda_0=0$. This will be assumed in further analysis.
Such a substitution leaves unchanged the total local induction parameter
\begin{equation}
\Lambda=\log(R_0/\xi)=\log(R_0/\delta) +\tilde\Lambda_0,
\end{equation}
where $\tilde\Lambda_0=\log(\delta/\xi)$. This property of system (4) makes
it possible to perform computer simulation with
smaller arrays of discrete points than that required
with small values of $\xi/R_0$, if we take $\delta$ larger than a few $\xi$. 
In the numerical experiments described below, the
main parameters are $\Lambda$ and $\delta$ (we usually set $\delta/R_0=0.05$
 and sometimes $\delta/R_0=0.025$ for comparison). We
use dimensionless variables so that $\Gamma=2\pi$ and $R_0=1$.
Coefficient $C$ in functions $\mu_j$ is chosen not too large to
comply with the numerical stability condition.

It should be noted that the system under investigation obeys the standard 
conservation laws for the Hamiltonian ${\cal H}$ (energy), momentum ${\bf P}$, 
and angular momentum ${\bf M}$:
\begin{eqnarray}
&&{\cal H}=\frac{1}{4}\sum_{j}\sum_{i}\oint\oint
\frac{({\bf X}'_j\cdot \tilde{\bf X}'_i)d\beta d\tilde\beta}
{\sqrt{|{\bf X}_j-\tilde{\bf X}_i|^2+\delta^2}}\nonumber\\
&&\qquad+\frac{\tilde\Lambda_0}{2}\sum_{j}\oint |{\bf X}'_j|d\beta,\\
\label{H}
&&{\bf P}=\frac{1}{2}\sum_{j}\oint[{\bf X}_j\times{\bf X}'_j]d\beta,\\
\label{P}
&&{\bf M}=-\frac{1}{2}\sum_{j}\oint|{\bf X}_j|^2{\bf X}'_j d\beta,
\label{M}
\end{eqnarray}
and the equations of motion taking into account the
freedom of parameterization possess the noncanonical
Hamiltonian structure $[{\bf X}'_j\times\dot{\bf X}_j]=\delta{\cal H}/\delta{\bf X}_j$.
The dynamic of torus vortices can be represented in the
canonical form by passing to the cylindrical coordinates 
and introducing $n$ pairs of $2\pi p$-periodic (in azimuthal angle $\varphi$) functions
$Z_j(\varphi,t)$ and $S_j(\varphi,t)=R^2_j(\varphi,t)/2$, which describe the 
shape of vortex filaments. The corresponding substitution must also be performed in
the Hamiltonian (8), which gives a rather cumbersome
expression that is not presented here. Significantly, the
equations of motion in this case have the canonical form
\begin{equation}
\dot Z_j=\delta{\cal H}/\delta S_j,\qquad -\dot S_j=\delta{\cal H}/\delta Z_j.
\end{equation}
The Hamiltonian description may turn out to be helpful 
in future analytic investigation of torus vortices, 
in particular, for theoretical interpretation of the
numerical results obtained below.

In this study, a pseudo-spectral scheme in variable $\beta$ 
and the fourth-order Runge-Kutta scheme
for integration with respect to time are used for
numerical simulation. The shape of each filament is
approximated by $L$ points ${\bf X}_{j,l}(t)={\bf X}_j(2\pi l/L,t)$ 
(typical values of $L$ are 512 and 1024). Here,
\begin{equation}
{\bf X}_{j,l}=\mbox{Re}\sum_{k=0}^{K-1}\hat{\bf X}_{j,k}\exp(2\pi {\rm i} kl/L).
\end{equation}
At each time step in the Runge-Kutta procedure, $K\approx(3/8)L$ 
corresponding Fourier harmonics are involved, after which only harmonics 
not senior to $K_{\rm eff}\approx L/4$ are left, while the remaining harmonics are
nullified. Such a technique proved to be quite effective
in various problems. In our case, it also demonstrates
a high stability and makes it possible to conserve integrals of motion 
${\cal H}$, ${\bf P}$, and ${\bf M}$ to within 5-7 decimal
places over the major part of the evolution (and to the
very end of run in the quasi-stability regions).

The time advancement is terminated when the
deformation of filaments becomes large enough or when a
certain large value of time $T_{\rm max}$ is attained (typically,
$T_{\rm max}=80$, but $T_{\rm max}=320$ is set in additional qualifying
experiments). For each set of parameters, final time $T_{\rm final}$
is recorded. In our numerical experiments, the measure 
of deformation of curves was the maximal value of
several harmonics with numbers close to $K_{\rm eff}$ (as a rule,
the increase in far harmonics indicated the convergence
of some segments of filaments and approach of the
reconnection instant). Time $T_{\rm final}$ generally depends on
the type of functions $\mu_j$ used; however, calculations have
shown that the exact form of the criterion for completion 
of an individual run is not very important for determining 
quasi-stable configurations; for this reason, all
details are not discussed here.

Since the symmetry of possible unstable modes
should not necessarily coincide with the initial symmetry 
of torus vortices, the initial conditions had to be
supplemented with perturbations containing ``nuclei''
of asymmetric modes. For $n=1$, the symmetry was
eliminated by multiplying the right-hand sides of
expressions (1) and (2) by $(1+\epsilon)$ and $(1+\epsilon)^{-1}$,
respectively (where $\epsilon\sim 0.01$) and by using parameter $\Theta_0$
incommensurate with $2\pi$; in some cases (for even $q$), 
the first term in expressions (3) had to be multiplied
by $1+0.005\sin(p\beta)$ for this purpose. For $n\ge 2$,
it was sufficient for symmetry breaking to slightly displace 
one of the vortices in the $(x,y)$ plane.

\begin{figure}
\begin{center}
\epsfig{file=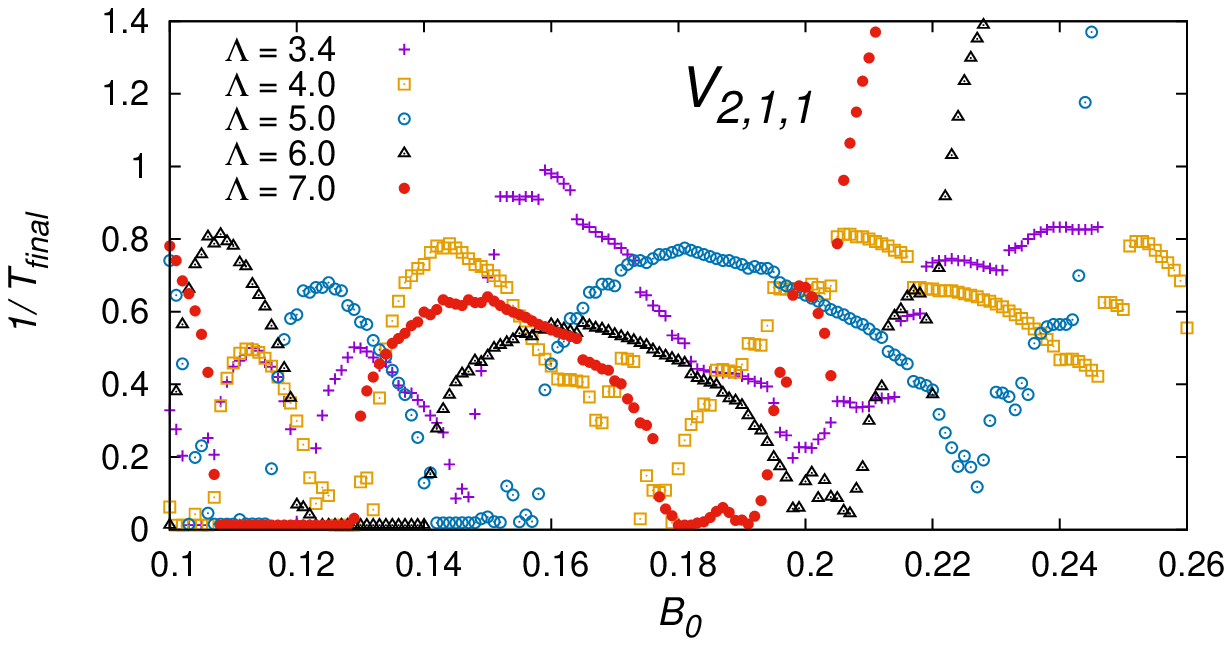, width=82mm}\\
\epsfig{file=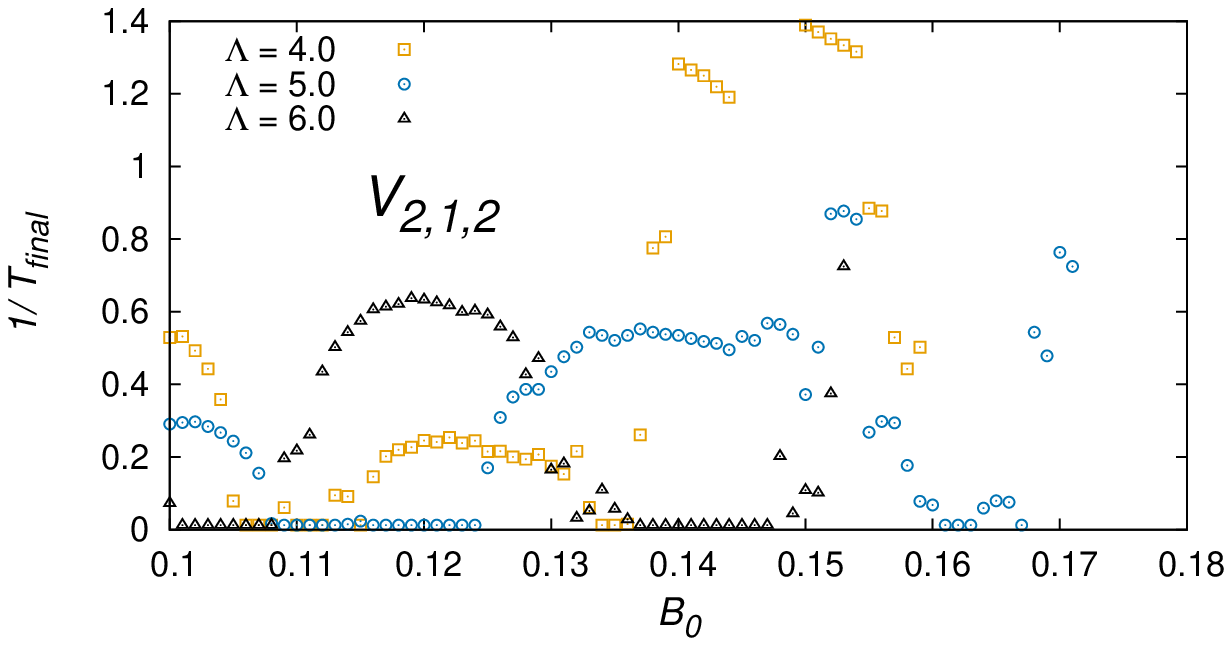, width=82mm}
\end{center}
\caption{Reciprocal lifetime for links $V_{2,1,1}$ and $V_{2,1,2}$.}
\label{V2-1q} 
\end{figure}
\begin{figure}
\begin{center}
\epsfig{file=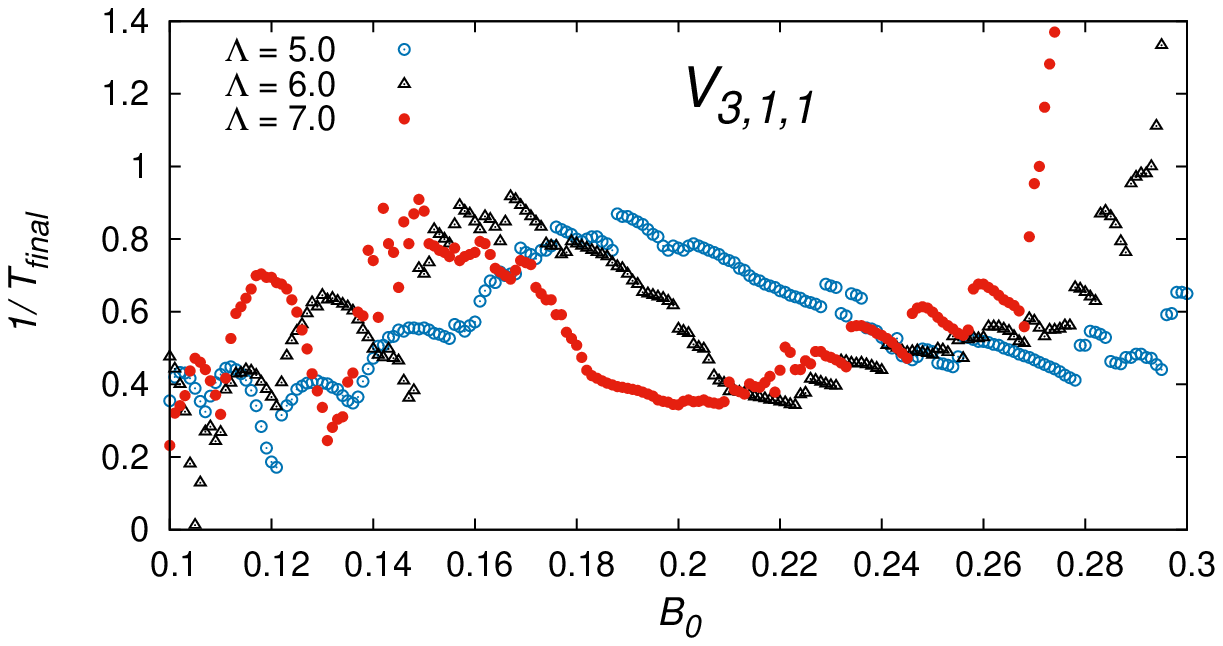, width=82mm}\\
\epsfig{file=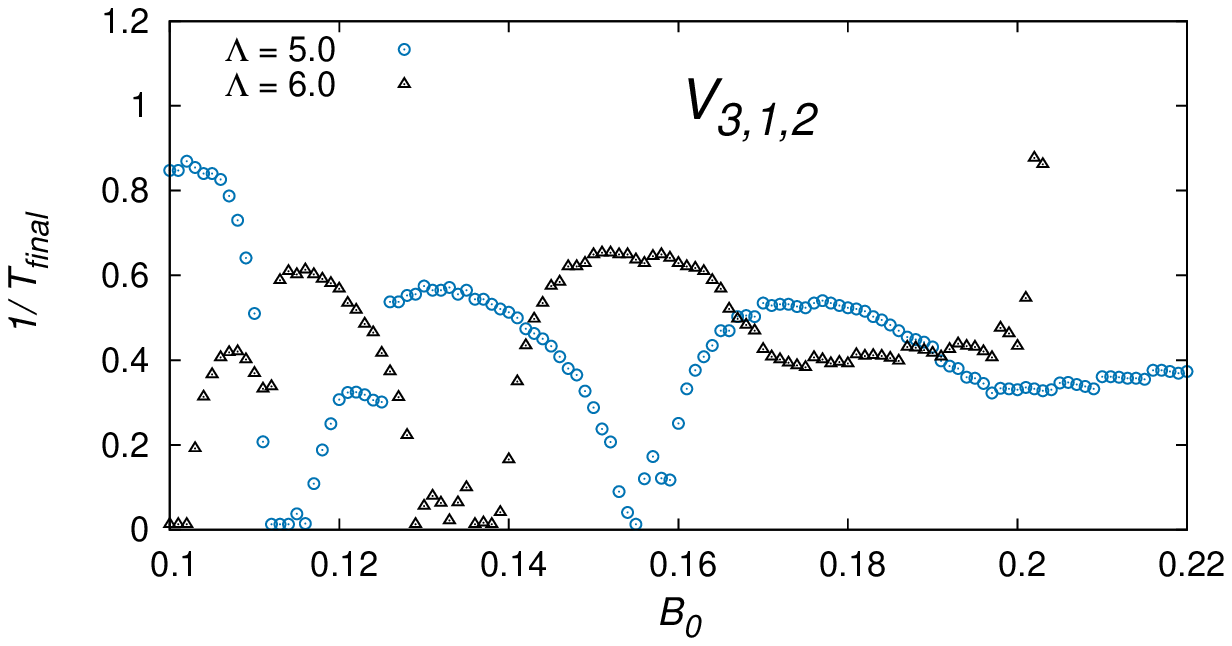, width=82mm}
\end{center}
\caption{Reciprocal lifetime for links $V_{3,1,1}$ and $V_{3,1,2}$.}
\label{V3-1q} 
\end{figure}

\section{Numerical results}

The numerically determined dependences of the
(reciprocal) lifetime on parameter $B_0$ for fixed values
of $\Lambda$  for different knots, unknots, and links are
shown in Figs. 1-7. Let us comment on these figures,
which are, in fact, the main results of this study.

It should be observed above all that a certain irregularity in the positions 
of points on the graphs (especially for not very small values of $B_0$) 
can apparently be explained by the noticeable difference between the 
initial conditions and the steady-state configurations.

It should be emphasized that the calculated reciprocal 
lifetime is not the instability increment because
the initial and final amplitudes of unstable modes have
not been fixed; actually, such modes themselves are
unknown. The lifetime generally depends on the
choice of the initial perturbations of the shape of torus
vortices. Only those (most interesting) segments of the
graphs, in which the points are close to the horizontal
axis, are more or less independent of this choice.

Typical values of $1/T_{\rm final}$ in all cases are on the order
of unity. However, some graphs contain small but
finite segments on which the reciprocal lifetime does
not exceed the small value of $1/T_{\rm max}$. In the vicinity of
the edges, the reciprocal lifetime exhibits an approximately 
root dependence. These are precisely the sought quasi-stability zones. 
These zones usually appear upon an increase in $\Lambda$, 
when adjacent parametric resonances of the dynamic system stop overlapping. 
Such a mechanism of the emergence of quasi-stable zones 
can be seen most clearly for knot ${\cal T}_{3,2}$,
unknot ${\cal U}_{2,1}$, and the simplest link of two rings  $V_{2,1,1}$. 
For many other $V_{n,p,q}$, the quasi-stability windows
do not appear at all. In particular, ${\cal T}_{4,q}$, and ${\cal T}_{5,q}$,
unknots ${\cal U}_{p\ge3,1}$, and three rings in configuration $V_{3,1,1}$
are unstable almost in all cases.

The upper panel in Fig. 1 shows that for a fixed
value of $\Lambda$, the difference in the results corresponding
to different values of $\delta$ is quite small in accordance
with the above remark. However, the difference still
exists because short-wavelength excitations of the
shape of a filament, which undoubtedly ``feel'' the difference 
between different values of $\delta$, appear due to
nonlinear interactions over long time intervals. For
this reason, the maximally accurate simulation based
on Eqs. (4) apparently still requires the use of original
(non-redefined) parameters $\xi$ and $\Lambda_0$. This
remark does not contradict the main conclusion concerning 
the existence of quasi-stable regions on the
plane of parameters $(B_0,\Lambda)$.

It is interesting to note that in some cases, quasi-stable
zones are located at large values of $B_0=$ 0.16--0.20, 
for which the initial torus cannot be treated as
thin any longer. In this case, the difference between
the initial conditions and steady-state solutions turns
out to be so large that we can speak on the preservation
of the shape of vortices only on the average. In fact,
vortex filaments oscillate quite strongly in the nonlinear 
regime; however, these oscillations astonishingly
do not cause the degradation of the system for a long
time. For example, in some auxiliary calculations,
vortex knots and links with parameters from such
stability zones propagated without noticeable
changes over distances exceeding thousands of initial
radii $R_0$. The existence of such solutions appears as the
most nontrivial result of this study.

\section{Conclusions}

Thus, the numerical experiments described above
revealed the existence of quite long-lived configurations 
for some types of torus vortex knots and links.
These results undoubtedly extended our knowledge
concerning a ``venerable'' fluid-dynamics problem. In
addition, these results are aesthetically attractive. At
the same time, a large number of new questions appear
because a rigorous theoretical description of this phenomenon 
does not exist at this stage. It is still unclear
whether it would be possible in the nearest future to
prepare and observe such quasi-stable quantum vortex
structures experimentally.

\end{document}